# Bottom-up Integration of TMDCs with Pre-Patterned Device Architectures via Transfer-free Chemical Vapor Deposition

Lucas M. Sassi,[1, #] Sathvik Ajay Iyengar,[1, #] Anand B. Puthirath[1, *], Yuefei Huang,[1] Xingfu Li,[1] Tanguy Terlier,[2] Ali Mojibpour,[3] Ana Paula C. Teixeira,[1,4] Palash Bharadwaj,[3] Chandra Sekhar Tiwary,[5] Robert Vajtai,[1] Saikat Talapatra,[6] Boris Yakobson[1,] and Pulickel M. Ajayan[1, *]

**Affiliations**

1. Department of Materials Science & NanoEngineering, Rice University, 6100 Main Street, Houston TX – 77005, United States.

2. SIMS laboratory, Shared Equipment Authority, Rice University, 6100 Main Street, Houston, TX – 77005, United States.

3. Department of Electrical and Computer Engineering, Rice University, 6100 Main Street, Houston, TX – 77005, United States.

4. Department of Chemistry, Universidade Federal de Minas Gerais, UFMG, Belo Horizonte, MG, 31270-901, Brazil

5. Metallurgical and Materials Engineering, Indian Institute of Technology Kharagpur, Kharagpur 382355, India.

6. School of Physics and Applied Physics, Southern Illinois University Carbondale, IL – 62901, United States.

Correspondence to: ajayan@rice.edu  (P.M.A), anandputhirath@rice.edu  (A.B.P)

# - These authors contributed equally to this work

**Abstract**

Two-dimensional (2D) transition metal dichalcogenides (TMDCs) remain a topic of immense interest. Specifically, given their low operational switching costs, they find many niche applications in new computing architectures with the promise of continued miniaturization. However, challenges lie in Back End of Line (BEOL) integration temperature and time compliance regarding current requirements for crystal growth. Additionally, deleterious and time-consuming transfer processes and multiple steps involved in channel/contact engineering can cripple device performance. This work demonstrates kinetics-governed in-situ growth regimes (surface or edge growth from gold) of $WSe_2$ and provides a mechanistic understanding of these regimes via energetics across various material interfaces. As a proof-of-concept, field effect transistors (FET) with an in-situ grown $WSe_2$ channel across Au contacts are fabricated, demonstrating a 2D semiconductor transistor via a "transfer-free" method within the 450-600 C 2h-time window requirement BEOL integration. We leverage directional edge growth to fabricate contacts with robust thickness-dependent Schottky-to-Ohmic behavior. By transitioning between Au and $SiO_2$ growth substrates in situ, this work achieves strain-induced subthreshold swing of ~140 mV/decade, relatively high mobility of $107 \pm 19$ $cm^2V^{-1}s^{-1}$, and robust ON/OFF ratios ~$10^6$ in the fabricated FETs.

**Key Words**

**Introduction**

Scaling silicon-based complementary metal-oxide-semiconductor (CMOS) technology has crossed the sub-10-nm threshold with recent demonstrations of 3 nm thick Si nanosheets gate-all-around transistor architectures poised to succeed FinFET structures beyond the 5 nm technology node.[1–3] However, further scaling becomes increasingly challenging due to gate electrostatics of the devices requiring a substantial reduction in the channel thickness to preserve desired performance.[2] In this context, two-dimensional (2D) semiconductors, such as transition metal dichalcogenides (TMDs), rise as a promising class of materials for the next generation of computing devices. These materials can maintain reasonable carrier mobilities even for stacks of atomic layers below 1 nm, which is not possible for thin three-dimensional (3D) semiconductors (such as silicon or germanium) as their carrier mobility degrades substantially for thicknesses below 4 nm, mainly due to increased scattering of charge carriers at the channel-to-dielectric interfaces.[1–3] Further, 2D semiconducting TMDs exhibit low-power switching and thus are ideal candidates in neuromorphic computing[4], sensor arrays[5], and low subthreshold swing FETs[6,7]. To realize these technologies[8], one of the most widely proposed pathways for 2D materials integration involves hybrid complementary metal-semiconductor-oxide (CMOS) strategies at the Back End Of Line (BEOL). Typically, BEOL integration aims to supplement and satisfy unique functionalities where traditional semiconductors fall short and is the likelier outcome when compared to Front End of Line integration for 2D technologies based on recent trajectories[9].

However, BEOL integration of 2D semiconductors has been challenging due to a strict thermal budget that cannot exceed temperatures of 450-600 °C within a 2 h processing limit[3,10,11]. Transfer methods have been attempted for low-temperature integration; however, these methods often introduce cracks, wrinkles, and polymer residue contamination, among other issues, substantially diminishing the mobility of the material and the performance of the device.[12]

Therefore, to use 2D semiconductors in Si-CMOS hybrid applications, developing a direct-transfer-free growth method as a bottom-up approach to enable area-selective growth of 2D layers on a CMOS-ready substrate is highly desirable.[1]

Many approaches have been developed to synthesize TMDs, such as mechanical exfoliation[13,14], liquid exfoliation[15], chemical vapor deposition (CVD)[16,17], metal-organic chemical vapor deposition[18], and molecular beam epitaxy[19]. CVD is one of the most promising, scalable, and in-budget synthesis techniques to combine the synthesis of large-area and monolayer TMD crystals[20]. However, typical CVD temperature synthesis for these crystals ranges from 700-900 °C,[21] which hinders direct growth compatible with the thermal budget of the BEOL process. In addition, spatial control of TMD growth is important for efficient implementation.[1] Strategies for patterned CVD growth involve using nucleation promoters, acidic salts, or metal seeds to reduce the energy barrier for the nucleation of the 2D crystals and enable location-selective growth.[22–24] However, the values of mobility measured with the resultant 2D crystals are very low ( < 35 $cm^2V^{-1}s^{-1}$). More recently, improved spatial control with a finite Schottky barrier in a non-epitaxial growth has been demonstrated, although, a direct-transfer-free growth method that yields relatively high mobility while maintaining Ohmic junctions is still to be demonstrated. To expand upon such in-situ growth techniques and reduce contact resistance, we seek inspiration from what is demonstrated as "hybrid" type contacts that involves significant channel material overlap over metal contacts[25] across a curved/recessed junction interfaces thereby improving transport properties[26].

In this work, we describe a bottom-up and one-step (transfer-free) CVD method to integrate the growth of TMDs with pre-patterned devices, which allows the precise deposition of TMDs onto desired locations guided by metal electrodes. All growth experiments were performed under BEOL-friendly conditions of 600 °C for under 30 minutes. We chose e-beam evaporated gold (Au) as a model for the metal contacts due to its wide and versatile use in literature and $WSe_2$ as our TMD model. The temperature required for the sublimation of $WO_3$ precursor is higher than that of $MoO_3$, and the temperatures required for selenization are generally higher than those required for sulfurization,[21], therefore, justifying that $WSe_2$ growth conditions would typically be the upper limit among TMD systems. Here, ohmic contacts with FET metrics such as $I_{ON}/I_{OFF}$, subthreshold swing, and mobility under strain show performance levels that overcome those reported for TMDs synthesized at high temperatures.

**Growth of $WSe_2$ on Au substrates and patterned structures**

We performed salt-assisted CVD[21,27] growth of $WSe_2$ on Au surfaces using $WO_3$ and Se as precursors and NaCl as a growth promoter on e-beam evaporated Ti/Au (5/300 nm) patterns made by photolithography at 600 °C (see Methods for details). The process schematics of the direct CVD growth of TMDCs on the Au patterns are shown in **Fig.1**. In stark contrast to previous observations of large triangular crystals obtained on Au foils[28–30], growth of $WSe_2$ occurs over the entire surface of the deposited gold film as shown in **Fig. 1a (i-iii)**. The large-area growth of $WSe_2$ on Au substrates is evidenced by Raman spectroscopy, as shown in **Fig. S1a**. We can see that the $E_{2g}$ and $A_{1g}$ bands (typically unresolved within 248 $cm^{-1}$)[31] are measured in 10 random positions across the Au film, demonstrating the growth of millimeter-scale $WSe_2$. In addition, the synthesized $WSe_2$ was also transferred to a $Si/SiO_2$ substrate for better optical contrast and visualization (**Fig. 1a-iii** and **Fig. S1b-ii**), therefore providing visual evidence of the growth of the millimeter/centimeter-scale $WSe_2$ film.

Moreover, we also observe that changes in the growth parameters, such as temperature, time, and precursor concentration, among others, may result in the growth of $WSe_2$ from the Au to the $SiO_2$ substrate when the growth is performed on patterned substrates, as shown in **Figs. 1b-i and S2a-c**. Furthermore, it is important to note that the growth from Au towards the substrate occurs in a stepped fashion, as shown in **Figs 1b-ii, 1b-iii, and S2a-c**, starting as a multilayer at the Au interface and gradually reducing in thickness until the $WSe_2$ is one layer thick. These steps can be several microns wide, as shown in **Fig S3**. This growth behavior indicates that the rate of growth of $WSe_2$ on Au is greater than on $SiO_2$. Therefore, by controlling the growth parameters, it is possible to adjust the rates of $WSe_2$ growth on the two substrates to synthesize $WSe_2$ structures with arbitrary shapes defined by a lithography process. Au square patterns have also been used to synthesize $WSe_2$ **(Figs. S2d-g)**. Raman spectroscopy mapping performed on the squares show that the $WSe_2$ is present in the whole square surface but not on the $SiO_2$ substrate, which shows that Au can be used to control the shape of TMDC flakes **(Fig. S2g)**. On the other hand, we also show that increasing the growth duration from 5 to 10 minutes promotes growth from beyond the Au squares **(Fig. S2f)**. In addition, increasing the growth temperature from 600 °C to 700 °C but keeping the time at 5 minutes also results in growth beyond the Au structures, as shown for the number markers in **Fig. S2a-c.**

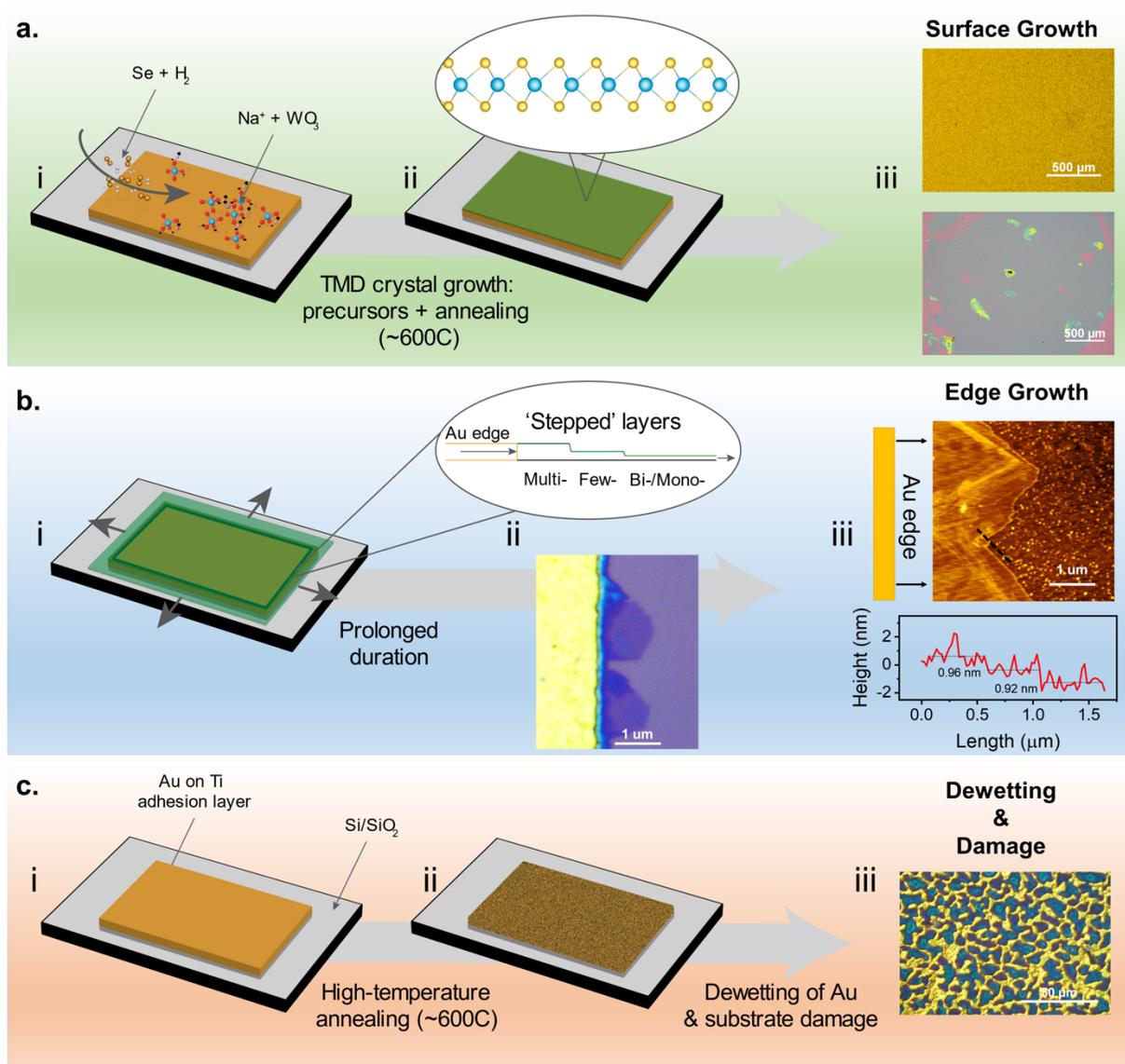

**Figure 1: Growth modalities on Au surface.** (a) Direct surface growth modality; (i) Introducing gas-phase precursors in a typical CVD process onto metal-substrate patterns, (ii) surface growth of TMDC layers with time-tunable thickness, (iii) optical micrographs depicting surface growth TMDC coverage (top) and corresponding TMDC coverage once transferred (bottom). (c) Prolonged growth duration yields edge growth modality; (i and ii) schematic illustration of an edge 'stepped' growth upon surface saturation and coverage creeping perpendicularly outwards, and optical microscopy images of optimized edge growth conditions (iii) AFM micrograph obtained from metal pattern edge exhibiting stepped behavior until monolayer edge on the substrate, (c) Heat treatment of typical metal-substrate patterns; (i) Au film with titanium adhesion layer atop Si/SiO$_2$ substrate, (ii and iii) heat treatment in the range of CVD TMDC growth requirements and optical micrograph depicting complete dewet, diffusion, and damage of metal-substrate interface.

Finally, to check the integrity of the Au structures during our growth process, we subjected the Ti/Au films to the CVD process without growth precursors (**Fig. 1c (i-iii)**). It can be seen in the optical microscopy image in **Fig. 1c-iii** that dewetting of the deposited Au occurs on top of the $SiO_2$ substrate. Since solid-state dewetting is driven by surface energy minimization, chemical species that are adsorbed/absorbed on the surface of the metal might change the surface properties and, hence, the dewetting behavior [32,33]. Therefore, it can be concluded that the sublimation of the precursors to the Au surface and the reaction of $WSe_2$ synthesis can protect Au against the dewetting process and prevent structural damage.

**Mechanism of $WSe_2$ growth on Au**

To understand the growth mechanism, the catalytic effect of Au, and the protective effect of the growth on the Au dewetting, we performed Time-of-Flight Secondary Ion Mass Spectrometry (ToF-SIMS) and X-ray photoelectron spectroscopy (XPS) on the samples of $WSe_2$ grown on Au films. Firstly, ToF-SIMS was used to investigate depth-dependent chemical distribution in the $SiO_2$/Ti/Au stack with respect to the presence and absence of growth precursors. In **Fig. 2a**, the depth profile describes the variation of the chemical composition from the surface of the Au (considered as 0 nm on the x-axis) to the interface between the Ti adhesion layer and the SiO$_2$ (set as 300 nm deep). In the absence of precursors (left), the depth profile shows that ion fragments of elements of interest, such as titanium (CsTi$^+$ and TiO$^+$) and oxygen (Cs$_2$O$^+$), diffuse toward the Au (Cs$_2$Au$^+$) surface leading to completed diffused interfaces and dewetting of the Ti/Au film. On the other hand, when the CVD process is conducted in the presence of precursors, the deposited stack preserves its interfaces with very limited diffusion of other elements to the surface of Au up to the first 100 nm from the surface. Any detected fragments' intensity is far below 10$^{-3}$ units on a log scale.

To investigate the top of the surface (first few nanometers) of the Au films, we performed a high-resolution depth profile using ToF-SIMS by reducing the Cesium sputtering to 500eV (10nA). **Fig 2b (i)** shows that both precursors used for the WSe$_2$ growth, WO$_3$ (CsWO$_3^+$), and Se (Cs$_2$Se$^+$), can be found at the first few nanometers inside the Au surface. However, the concentration of the precursors becomes significantly reduced when the measurement gets further from the Au surface. This result is also in agreement with our XPS observations. **Fig. 2c (i)** shows the high-resolution spectrum of the W 4f state of the WSe$_2$ grown on Au, in which we can observe two sets of spin-orbit split doublets with a separation of 2.17 eV. The first has its peaks around 32.7 eV and 34.9 eV corresponding, respectively, to the 4f$_{7/2}$ and 4f$_{5/2}$ lines of W$^{4+}$, which are ascribed to WSe$_2$ and the second doublet has its peaks around 36.0 eV and 38.2 eV, which corresponds to the 4f$_{7/2}$ and 4f$_{5/2}$ lines of W$^{6+}$ and W$^{5+}$ chemical states and are assigned to the formation of WO$_x$ (x ≤ 3), therefore indicating the presence of tungsten oxides within the Au surface.[34], which can appear due to an incomplete reaction of WO3 with Selenium in a slightly deeper region of the surface of the substrate.

Regarding the Se 3d core level spectrum in **Fig. 2c-ii**, we note that it exhibits the doublet with peaks at 54.9 eV and 55.8 eV (separation of 0.86 eV), which are assigned to the 3d$_{5/2}$ and $3d_{3/2}$ lines of the $Se^{2-}$ chemical state in the $WSe_2$[34]. However, we can also note 3 more peaks in the XPS spectra not commonly found in the Se 3d spectrum of $WSe_2$ grown on $Si/SiO_2$ substrates. The peak at 57.9 eV can be ascribed to the Au $5p_{3/2}$, which is not present in pure Au but appears due to the interaction of Se and Au[35]. In addition, the peak at around 61.0 eV is associated with the formation of selenium oxide[36,37], which can occur due to an incomplete reaction of Se with the precursor $WO_3$ or to the formation of tungsten oxyselenide[38]. Finally, the last extra peak at 63.8 eV can be assigned to the Na 2s state[37]. The presence of Na shows residual traces of precursor at the surface of Au as the NaCl, which is used as a growth promoter, reacts with the $WO_3$ in high temperatures to form $Na_2WO_4$, which, in turn, is one of the main compounds that participate in the synthesis reaction of $WSe_2$[39,40]. It is important to point out that the penetration depth of XPS is at the order of 3-10 nm. Therefore, it is likely that part of the precursors detected with XPS, such as tungsten oxides, Se and Na, is not exactly on the surface of Au but diffuses a few nanometers inside the Au film, in agreement with the ToF-SIMS measurements in **Fig. 2b (i)**. In addition, we calculated the adsorption energies of the precursors Se and $WO_3$ with respect to Au substrate. The results, shown in Fig 2b (ii), help to explain the excess of Se compared to $WO_3$ as the adsorption energy for Se is considerably higher than the one for $WO_3$. This agrees with our measurement of the Se/W ratio of 3.67 from XPS and with the previous observation that more Se diffuses inside the Au film compared to the tungsten precursor (**Fig. 2b(i)**).

To understand the large differences in the rate of growth of $WSe_2$ on Au and $SiO_2$ observed in our experiments, we examine the affinity of different surfaces (Au, $SiO_2$, and pre-grown $WSe_2$) to $WSe_6$ molecules in the gas phase by calculating the adsorption energies from density functional theory (DFT) calculations. The salt-assisted CVD growth of a similar TMDC, $MoS_2$, proposes that the fully sulfurized $MoS_6$ molecule emerges as an intermediate gas precursor to the crystal growth[41,42]. The calculated energies for these three substrates are, respectively, -3.87 eV, -0.748 eV, and -1.16 eV, as shown in Fig. 2d. We note that the binding energy for the intermediate on Au is much greater compared to other substrates and that the binding energies for $SiO_2$ and $WSe_2$ are comparable. Therefore, this result confirms that Au is a more favorable substrate for the growth of $WSe_2$ and corroborates the fast rate of growth on Au that we observe experimentally.

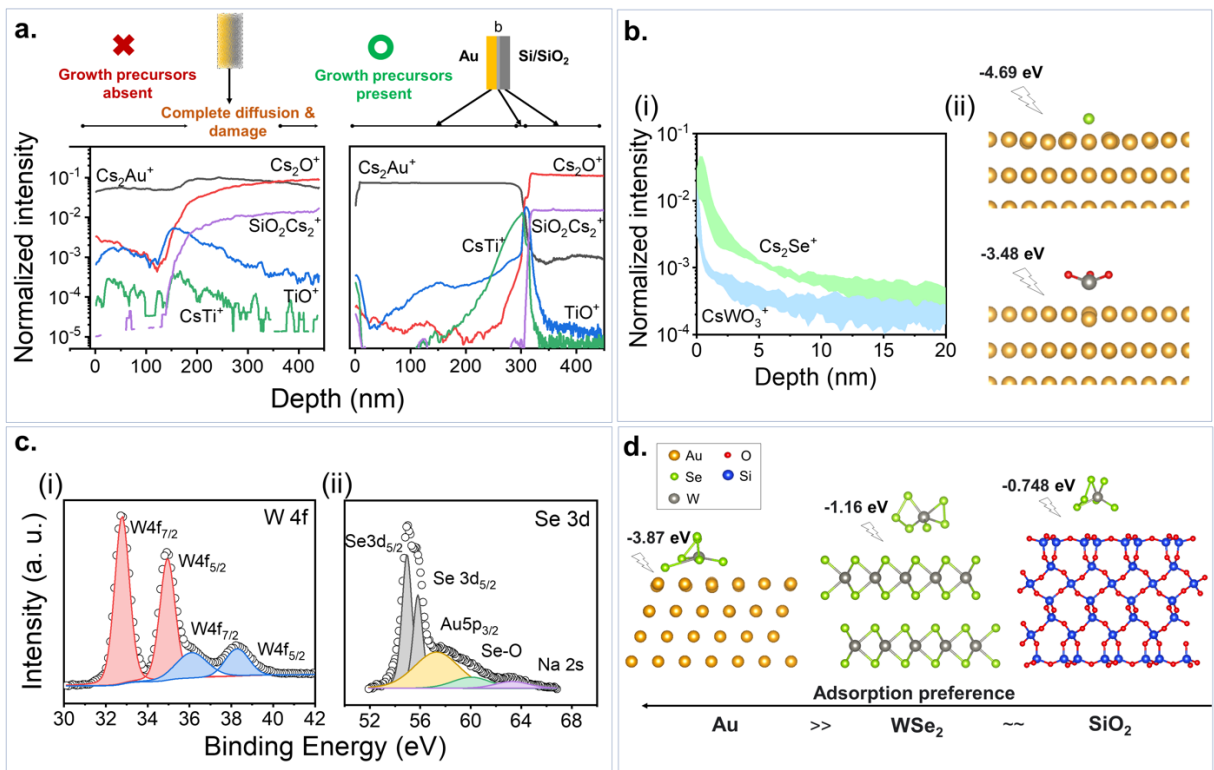

**Figure 2. Evidence for intact device components, post-CVD, and underlying growth mechanism.** (a) Time of flight secondary ion mass spectrometry (TOF-SIMS) depth profile in the absence and presence of growth precursors post-CVD conditions indicates intact substrate and compositional layers when precursors are present, (b) TOF-SIMS depth profile indicates significantly greater Se precursor diffusion and penetration into the Au surface than $WO_3$ and corresponding adsorption energies indicating a greater surface preference for Se, (c) X-ray photoelectron spectroscopy of Au surface exposed to precursors under CVD conditions before growth. Precursor presence is observed on the top surface of Au. (d) Comparison of adsorption energy across various growth surfaces during CVD: Au, pre-existing $WSe_2$ on Au, and $SiO_2$.

Observing precursors inside the Au structure by ToF-SIMS and XPS and that the $WSe_2$ grows from outside patterned gold structures by increasing the growth time might indicate a liquid-mediated synthesis process. Using liquid metal as a reaction environment was shown before for the growth of 2D oxides[43] and the growth of TMDCs nanoribbons using Au[44] or Ni particles as precursor[45]. These works show that precursors are found inside the catalyst particles and that the growth rate in a liquid-mediated process is much faster than in a vapor-solid-solid (VSS) process, at least 20 times higher. Although our process temperature (600 °C) is far below the melting point of Au (1064 °C), Liu *et al.*[46] showed, via molecular dynamics, that the grain boundaries in nanocrystalline Au get liquefied at temperatures far away from the corresponding melting point of Au. Therefore, a liquid-mediated synthesis process of growth occurring in the nanocrystalline structure of the e-beam evaporated Au would provide a feasible explanation for the high selectivity and preference of the $WSe_2$ growth on the Au structures. In addition, it would also explain the differences in the $WSe_2$ obtained from the evaporated Au structures with respect to the triangular crystals with limited domains obtained on Au foils[29].

**Transfer-free growth and evidence of strain**

Using the concept of different growth rates of $WSe_2$ on Au and $SiO_2$, $WSe_2$ was grown between Au electrodes to develop a transfer-free device fabrication process directly via CVD. As an extension of surface growth, stepped edge growth was utilized to bridge contact between two patterned Au electrodes. As shown in **Fig. 3a**, fixing all growth parameters (such as precursors concentration, substrate and precursor position, and flow rates) and tuning only growth time presents a simple time-dependent optimization to yield monolayer $WSe_2$ between the electrodes. This progression is captured in **Fig. 3a (i) to (ii)**; $WSe_2$ at the electrode interface is multilayer and thinner towards the center of the channel. Scanning electron microscopy (SEM) contrast in **Fig. 3a(iii)** indicates surface growth on Au (discoloration) along with edge growth across the junction and confirms the absence of Au dewetting. The atomic force microscopy (AFM) image can confirm the monolayer (measured thickness of 0.7-1 nm[25,26] at the midpoint in **Fig. 3b**. Macroscopic defects, such as wrinkles or cracks, are visibly absent.

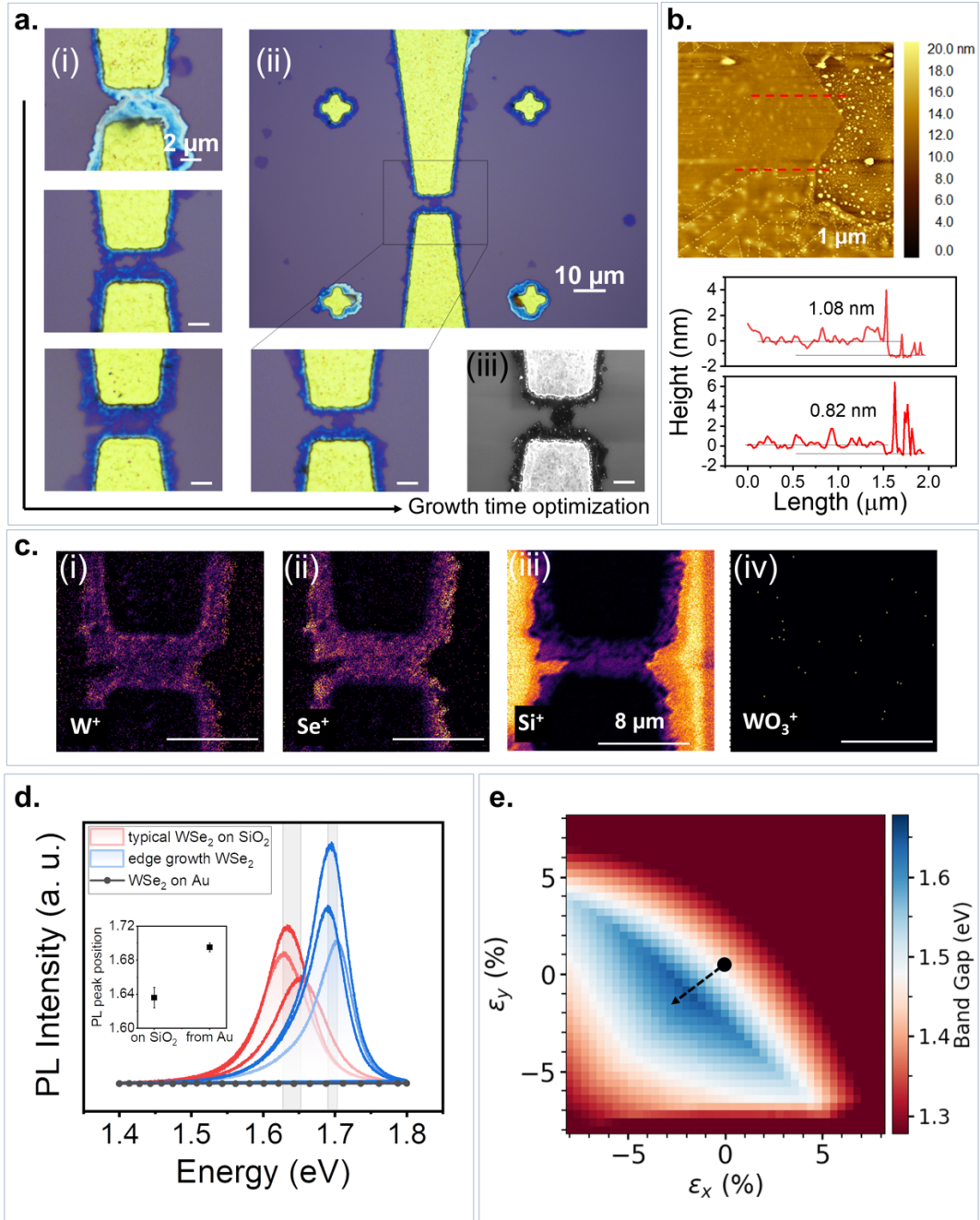

**Figure 3: Growth of Transfer-Free Junctions**. (a) Optimization of time for the growth of the direct-transfer-free device via CVD. (b) AFM image of the WSe$_2$ flake at the center of the junction and height profile along the marked dashed lines AFM image confirming monolayer junction WSe$_2$, (c) TOF-SIMS analysis of the WSe$_2$ at the center of the junction confirming continuity of material without unreacted precursor residues, (d) Photoluminescence spectra depicting intensity blue-shift for edge growth WSe$_2$ hinting at a strained system during the surface-to-edge transition, (e) Dependence of WSe$_2$ monolayer bandgap on strain states calculated from first principles. The black dashed line indicates that moving towards compressive strain yields a significant blue shift.

As the time-of-flight secondary ion mass spectrometry (ToF-SIMS) yields molecular information through ion fragments that are sputtered by the incoming $Cs^+$ ion beam, it is a more deterministic tool to distinguish between residual precursor ($WO_3^+$ from $WO_3$) and target TMDC material ($W^+$ and $Se^+$ from $WSe_2$) in comparison to XPS that relies on identifying very minute shifts in binding energies.[47] **Figs. 3c (i-iv)** shows the ToF-SIMS analysis of the region between the electrodes operating in surface mode depicting a uniform distribution of $W^+$ and $Se^+$ ions between the electrodes and $WO_3^+$ just at the edges of the flakes. $Si^+$ map highlighting the background of the device also supports the argument that the growth process has not damaged the substrate. The near absence of $WO_3$ altogether confirms the purity of our $WSe_2$ junctions. While $WO_3^+$ ions were not found on the top surface of the electrodes, XPS captured the presence of $WO_3$ in the Au due to a higher penetration depth (resolution 3-10 nm) than surface imaging modes via TOF-SIMS (resolution 1-3 nm), indicating that the residual precursor is in the deeper layers of Au. In contrast, only $WSe_2$ can be found at the surface.

Transitioning substrates from Au to $SiO_2$ during growth would warrant a strained $WSe_2$.[48] Photoluminescence spectra (PL) in **Fig. 3d** confirm this by a significant blue-shift (Δ=0.06 eV) Of the $WSe_2$ that grows from the gold on the $SiO_2$ with respect to the $WSe_2$ that grows directly on $SiO_2$. This observation holds even when the PL peaks are deconvolved to measure the blue shift in the A exciton peak while excluding the effects of the trion peak causing overall peak asymmetry (see **Fig S4a**). Typical PL quenching is also observed for $WSe_2$ directly grown on the surface of Au (Fig.S1d). This has been observed for TMDCs on metallic or carrier-rich surfaces due to spin-orbit and interlayer coupling.[49,50] It is important to point out that although the PL of a directly grown $WSe_2$ on Au is quenched, it is obtained when the sample is transferred to a dielectric substrate such as $WSe_2$. Moreover, when $WSe_2$ is grown on SiO2 and subsequently transferred to Au, the signal does not get quenched due to bad interlayer coupling, which indicates that growing the TMDC directly on the Au surfaces promotes a good coupling interaction between the two materials.

Interestingly, our first-principles calculations of bandgap shift as a function of strain for monolayer $WSe_2$ (**Fig. 3e**) reveal that the bandgap tends to blue-shift under low to moderate compressive strain and red-shift under extreme compressive and any form of tensile strain. This trend is highly consistent with recent reports involving compressive strain across $MoS_2$ (axial) and Janus S-W-Se (hydrostatic).[51,52] **Fig. S4b** captures a similar trend for bulk $WSe_2$, hinting at a low correlation to layer thickness due to weak interlayer coupling. Therefore, PL and first-principles calculations correlate the blue shift observed to compressive strain.

Moreover, while the trends observed are of greatest significance, the origin for a discrepancy in the value of baseline (under zero strain) bandgap value between that simulated (1.478 eV) and experiment (1.636 ± 0.012 eV, $Si/SiO_2$) can be ascribed to two main aspects. First, experimental demonstrations have revealed varying values for PL for $WSe_2$ based on varying strains across different growth substrates.[53] Second, DFT calculations with LDA or PBE-GGA exchange-correlation functional is known to result in a bandgap underestimation of approximately 30% for TMDCs.[54]

**Microscopic Evidence of Surface-to-Edge Growth Transition**

First, the quality and phase of $WSe_2$ are verified through atomic resolution Transmission Electron Microscopy (TEM). A cross-sectional lamella was prepared by Focused Ion Beam (FIB) milling to examine surface and edge growth regimes and the integrity of the in situ grown transistor under TEM. The bright field images of the prepared lamella with a thickness < 80 nm are depicted in (**Fig 4a**). **Fig 4b** shows that the $WSe_2$ grown on the surface of Au constitutes multilayer stacks. This crystal structure is consistent with lattice parameters for 2H semiconducting $WSe_2$. Moreover, images of the lamella at four different points: the electrode surface, at the edges of the electrodes, at the electrode-channel interface, and finally, at the channel region, captures the essence of the WSe$_2$ growth transition between Au and SiO$_2$ substrates. The thickness difference between the WSe$_2$ layer grown on the electrode surface and across the channel region confirms that surface growth precedes edge growth.

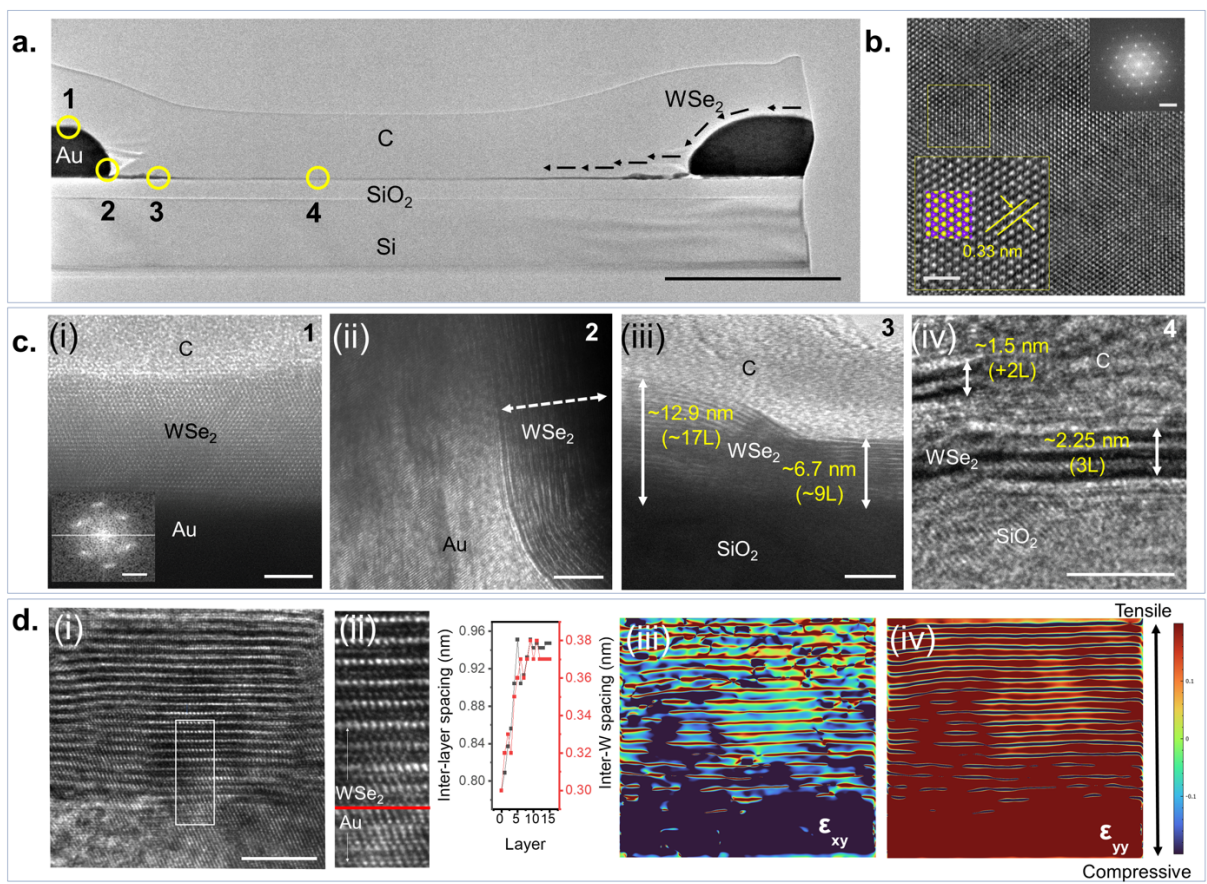

**Figure 4. Transmission electron microscopy of surface-to-edge growth transition.** (a) Cross-sectional TEM bright field image of Au-WSe$_2$-Au junction protected by a layer of graphitized carbon upon Si/SiO$_2$ highlighting 4 regions of interest. The scale bar is 2 um. (b) Atomic resolution TEM of WSe$_2$ cross-sectioned area, scale bar is 5 nm; insets indicate the zoom-in area with WSe$_2$ space-filling model overlap and W-W spacing of 0.33 nm (scale bar is 1 nm) and Fast Fourier Transform (FFT) diffraction pattern corresponding to the entire WSe$_2$ area shown (scale bar is 5 nm$^{-1}$). (c) High-magnification TEM images of regions of interest along junction cross-section reveal a gradual and curved transition from surface growth to edge growth, (i) multilayer WSe$_2$ with crystalline ordering and corresponding FFT diffraction pattern (scale bar is 5 nm$^{-1}$), (ii) transition from the surface (on Au) to edge growth (from Au and upon SiO$_2$) by strained bending of the multilayer stack, (iii) considerable reduction in layer thickness along the junction axis (smoother steps) until (iv) few-layer thick WSe$_2$ is observed (sharper steps). (d) (i) cross-sectional image of the WSe$_2$-Au interface, (ii) zoom-in of selected interface area depicting positive trends in interlayer spacing and inter-W spacing, (iii and iv) relative strain maps of interface based on lattice periodicity. All scale bars are 5 nm.

Moreover, under higher magnifications, it is observed that stacks of multilayer $WSe_2$ bend to accommodate this surface-to-edge growth regime transition (**Fig 4c(ii)**). TMDC layers grow conformally on curved surfaces minimizing residual strain in the films. This has been observed based on various interfaces formed by TMDCs core-shell structures. [55] Elevated temperatures during the growth process can also favor smoothening Au surfaces[56]. While this strain-induced $WSe_2$ bending continues across the $SiO_2$ surface and contributes to the curved "step"-like features (**Fig 4c(iii)**), the effects of strain, and therefore bending, are diminished far away from the Au surface, leading to sharper and thinner steps (**Fig 4c(iv)**). Darkfield imaging and elemental mapping reveal robust junction contacts without breakage or damage (**Fig S5 EDS and HAADF**). While the transition of $WSe_2$ from Au to $SiO_2$ is significant, the interface with Au is equally essential, and **Fig 4d (i and ii)** capture this interface with a zoomed-in view. Specifically, we note a clear transition from a strained lattice (compressive at the Au surface) to an unstrained lattice (far away from the surface) in the form of positive trends in interlayer spacing and inter-W distances with increasing layer counts. These measurements directly translate to strain extracted from the sample FFT (**Fig. S7**), visualized in **Fig 4d (iii and iv)**. We observe highly compressive strain at the Au surface, which morphs into unstrained conditions away from the Au surface, consistent with our previous strain-bandgap simulations.

**Transport Across In-situ Grown Devices**

One of the biggest challenges in designing 2D materials-based devices lies in engineering good contact between the metal electrodes and active channels while avoiding the deleterious effects of a transfer process. Additionally, each step in the device fabrication process can introduce microscopic and macroscopic defects. Incorporating this concept as a one-step chemical synthesis initiated from the metal pads leads to better contacts, as neither edge nor top contacts. Still, a third hybrid case with significant top and edge metal-semiconductor overlap is optimal[25]. This also leads to a gradual layer thickness variation in a stepped fashion in our growth time-optimized devices. The stepped behavior yields interesting trends in IV characteristics that serve as a basis for tuning junction contact quality.

As shown in **Fig. 5a**, A thick, heterogeneous multilayer junction unequivocally demonstrates a typical Schottky behavior owing to its low knee voltage (0.7 V) and broad reverse regime. However, as the junction begins to adopt a more stepped-like morphology and the resulting in-situ growth contact is thinner, this Schottky behavior diminishes greatly and approaches Ohmic characteristics. Not only is the stepped morphology thinner in the center of the channel, but as observed previously, the surface coverage of WSe$_2$ is also thinner. This is due to the successive nature of growth: surface followed by edge modalities. In the stepped-until monolayer case (**Figs. 1b (ii and iii) and 3a (ii)**), we observe a typical Ohmic response over a broad potential range, albeit with highly diminished residual Schottky features.

The transition from Schottky to Ohmic contact behaviors was not abrupt. Still, it was found to be strongly dependent on the number of layers of WSe$_2$ present in the conducting channel of the devices investigated. This observation establishes a gradual trend from one mode to the other. Thicker samples tend to have stronger nonlinearity, lower current at low bias, and higher current at high bias. This trend can be explained by a two-step transport model as shown in **Fig. 5b**: (1) carrier transport through the Au electrode to WSe$_2$, and (2) carrier transport in WSe$_2$. This can be written as:

$$V_{\text{eff}} = V_0 + V_{\text{x}} \quad \ldots (1)$$

The first step causes the nonlinearity in the I-V curves, which follows a typical transport equation across a metal-semiconductor interface:

$$I = I_0[exp(\frac{qV_0}{\eta kT}) - 1] \quad \ldots (2)$$

Equation (2) is commonly used to characterize the Schottky barrier height[57], where $\eta$ is the ideality factor. Due to bending of WSe$_2$ when extending from Au electrode to SiO$_2$ substrate, the first few layers of WSe$_2$ may terminate at SiO$_2$, instead of connecting to WSe$_2$ that grows on SiO$_2$. Carriers have to transport through more layers for thicker samples. Thus, thicker samples have a lower current in their I-V curve for this step. The second step is to transport in WSe$_2$, where I-V has a linear relation V = IR, with R smaller for thicker samples. The two steps are connected; thus, our non-trivial IV curves can be characterized by a combination of a logarithmic term (rearrangement of eq (2)) and a linear term as shown below (inclusive of Schottky, Ohmic as well as interfacial features):

$$V_{\text{eff}} = \eta kT \, log(\frac{I}{I_0} + 1) + IR \ldots (3)$$

Fitting experimental data to equation (2) (**Fig 5a** inset) yields a positive correlation between $\eta$ and sample thickness and a negative correlation between $R$ and sample thickness, consistent with previous studies[58]. The Schottky behavior of the first step is qualitatively shown in **Fig5b**, with a first principle 2-terminal electron transport calculation. An Au|WSe$_2$ interface is built by attaching monolayer WSe$_2$ to Au (111) surface. The calculation's left and right electrodes are Au and WSe$_2$, respectively. The contribution to IV curve behavior from the stepped morphology itself, is found to be negligible in our transport simulations (see **Fig. S7**).

To investigate the effect of carrier excitation on these devices, we performed temperature-dependent transport measurements considering the extreme cases in the tunability regime (**Fig 5c**). As our devices normally operated under ambient temperature conditions (300K) (see **Fig.S8** for ambient condition sample aging study), we studied IV characteristics under elevated temperatures up to 370K. The Schottky case contacts tend towards a more Ohmic behavior at higher temperatures indicating a lowering of Schottky barrier height– consistent with n-type solid systems.[59] In contrast, the Ohmic case contact further improved in quality. Back-gated transistor measurements across the Ohmic contact junction (**Fig 5d**) reveal a high threshold voltage, consistent with several reports on 2D TMDC devices[60,61]. **Fig 5e** compares output characteristics for a Schottky and an Ohmic case device, highlighting the presence and absence of linearity, respectively. Linearity is maintained across various gating voltages, demonstrating Ohmic stability. Several blank measurements reveal only typical magnitudes of leakage current (~$10^{-12}$ A/μm) (**Fig S9-10**). Further, as strain plays a significant role in enhancing mobility in 2D materials[62], the Ohmic junction is found to have a subthreshold swing of ~140 mV/decade and mobility of ~107 ± 19 $cm^2V^{-1}s^{-1}$, specifically among CVD-grown devices (**Fig 5f**). Lastly, corroborating with observations in cross-sectional microscopy about the geometry of the Au-WSe$_2$ interface, "recessed" or curved contacts have been demonstrated to possess superior transport properties[26], supporting our relatively higher mobility observations.

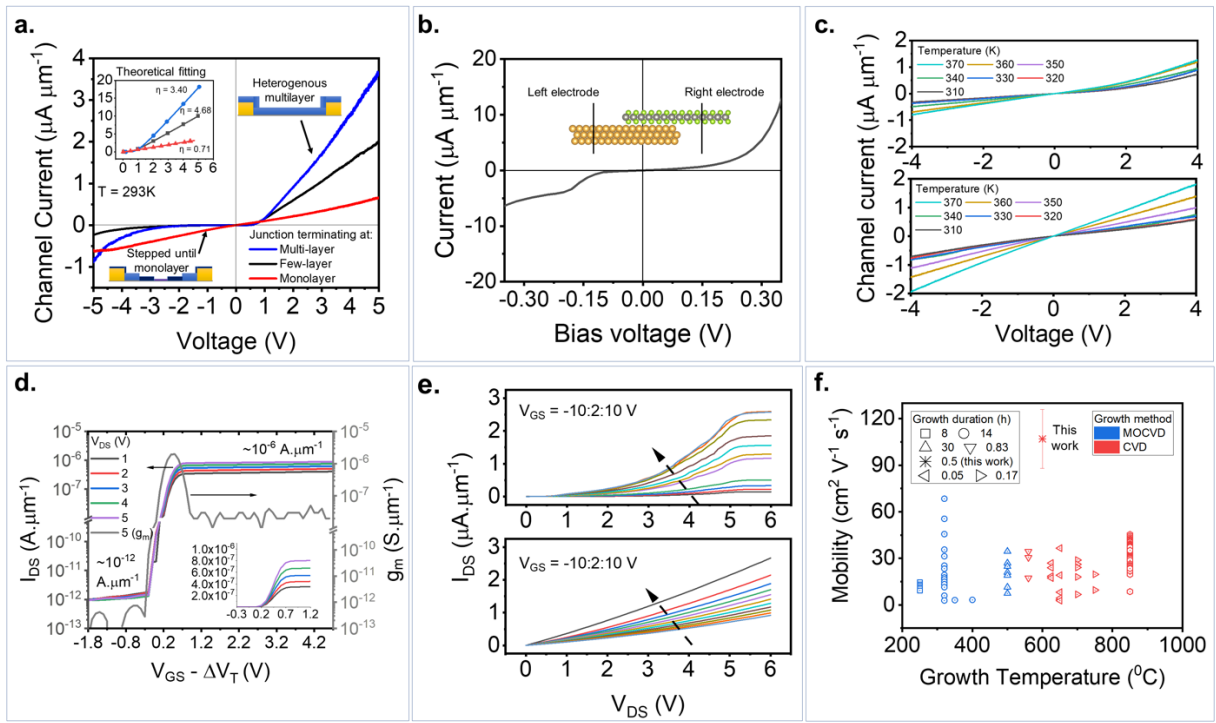

**Figure 5. Schottky-to-Ohmic trends and field effect transistor (FET) performance.** (a) 2-terminal measurements across heterogeneous multilayer (Schottky-like), stepped until monolayer (Ohmic-like) and stepped until few-layer cases indicating a trend in IV behaviors, (b) First principles calculation of I-V curve of Au|WSe$_2$ interface using DFT+NEGF methods, (c) Temperature-dependent measurements to observe IV curve stability reveal the absence of activation barriers, (d) Transfer characteristics of a back-gated FET prepared on a stepped until monolayer case device. Inset indicates zoom-in of field modulation. (e) Output characteristics of FET for a heterogeneous multilayer case (top) and a stepped-until monolayer case (bottom), (f) Mobility vs. processing temperature comparison among CVD/MOCVD-grown samples.

**Conclusion**

This work presents a functionally competitive route to prepare metal-TMDC junctions in situ without the deleterious TMDC transfer step. We shed light on the underlying mechanisms of WSe$_2$ growth on thin films of Au and how it contrasts with contemporary work on foils enabling large-area surface coverage and growth. Using an in-situ growth method, we demonstrate substrate protection and the absence of metal dewetting while simultaneously operating within the BEOL limitations. Using surface and edge growth modalities, we bridge metal contacts in a junction that is demonstrated to possess strain, thereby superior FET metrics compared to standard transfer-based junctions. We envision this methodology opening new avenues for hybrid fabrication approaches such that TMDCs can supplement where they cannot supplant.

**Materials and Methods**

**Au patterning and electrode fabrication**

Au patterns and electrodes were prepared using photolithography and e-beam evaporation. Si/SiO$_2$ wafer chips (p-doped 300 nm thermal oxide) of 1.2 x 1.2 cm size were diced and solvent cleaned (acetone, methanol, isopropanol). The spin coat recipe for Shipley s1813 positive photoresist and liftoff resist (LOR) was 3500 rpm for 45 s. First, the Shipley was spin-coated and baked at 180 C for 4 min, followed by a LOR spin coat and bake at 110 C for 60 s. Baked samples were then exposed to UV light utilizing an EVG 620 Mask Aligner. The samples were developed in MF319 for the 50s, washed in DI water for 20s, and dried. ~5nm Ti (adhesion layer) was evaporated onto the sample, followed by 300 nm of Au. PG Remover (an NMP organic solvent formulation) was used as the liftoff agent, resulting in patterned structures.

**In-situ transfer-free device fabrication**

Patterned Au electrodes on Si/SiO$_2$ were subjected to a chemical vapor deposition (CVD) process in the presence of 50 mg of anhydrous WO3 powder (powder, purity., 99.9%; Sigma-Aldrich) and 4 mg of NaCl salt mixed in a ceramic boat. The boat was then placed in the center of a 2-inch Lindberg Blue tube furnace with the substrates facing down towards the precursor powder blend. Roughly 5g of Se powder is upstream laced in a crucible. Under a flow rate of 100 sccm of Ar/H$_2$ (85% Ar, 15% H$_2$), the setup was subjected for 20 min to purge the tube at room temperature, followed by a ramp up to 600 $^0$C in 15 min and sustained for 20 min (optimized: 20min, undergrowth: 15min, overgrowth: 30min). The setup was then allowed to cool to 500 $^0$C, after which the furnace was opened to quench and cool the setup to ambient conditions. Once the sample is cooled, an indium bead is pressed onto the Si (wafer side), which serves as a back gate contact.

**Transfer for quality check**

The samples grown on Au or patterned Au structures on SiO$_2$ were transferred using a wet chemical transfer method. Poly (methyl methacrylate) (PMMA) was first spin-coated on the as-grown samples at 3000 rpm for 60 s. The samples were then submersed in a 1M KOH solution to etch the underlying SiO$_2$ layer. After removing the SiO$_2$ substrate, the samples were then transferred to a gold etchant solution (etching rate ~ 2Å/s), composed of a KI/I$_2$ solution, to remove the Au film. The floating PMMA containing the sample was washed multiple times with DI water to remove residues from the solvents and then scooped onto a fresh SiO$_2$ for further characterization. The sample was then immersed in an acetone bath overnight to remove the PMMA and then rinsed with acetone and isopropyl alcohol.

**Materials characterization**

Raman spectroscopy and photoluminescence measurements were performed at room temperature using a Renishaw inVia confocal microscope. The spectra were acquired using a 532 nm excitation laser focused via a 50x objective lens with a spot size of 1 µm, laser power of 20 µW and acquisition times of 10s. Atomic force microscopy images were obtained in tapping mode using a Park NX-10 microscope. The measurements were carried out in air using silicon nitride cantilevers at room temperature. XPS spectra were obtained using a PHI Quantera system with an Al Kα X-ray source (1486.7 eV, 50 W). The beam spot size used in the measurements was 200 µm, and the pass energy was 26 eV. After Shirley background subtraction, the high-resolution XPS spectra were fitted with Gaussian-Lorentzian functions. For the curve fitting, the following constraints were applied: The spin-orbit splitting for the W $4f_{7/2}$ and W $4f_{5/2}$ components were fixed at 2.17 eV, and the area ratio between these components was of 4:3, while the separation for the Se $3d_{5/2}$ and Se $3d_{3/2}$ was of 0.86 eV with an area ratio of 3:2. Scanning electron microscopy was performed using an FEI Quanta 400 operated at 10 kV. Transmission electron microscopy was used to analyze the material by preparing a thin cross-sectional lamella via a focused ion beam milling process utilizing a Helios Nano Lab 660 FIB unit, and the microscopy is performed with aberration-corrected Titan Transmission electron microscope (FEI Titan Themis 3) with an acceleration voltage of 300 kV.

**Time-of-flight secondary ion mass spectrometry**

ToF-SIMS measurments were performed using a ToF-SIMS NCS instrument, which combines a TOF.SIMS instrument (ION-TOF GmbH, Münster, Germany) and an *in-situ* Scanning Probe Microscope (NanoScan, Switzerland) at the Shared Equipment Authority of Rice University. The analysis field of view was 80 × 80 µm$^2$ ($Bi^{3+}$ @ 30 keV, 0.35 pA) with a raster of 128 x 128 along with the depth profile. A charge compensation with an electron flood gun has been applied during the analysis. An adjustment of the charge effects has been operated using a surface potential. The cycle times was fixed to 90 µs (corresponding to m/z = 0– 735 a.m.u. mass range). The sputtering raster was 300 × 300 µm$^2$ (Cs$^+$ @ 2 keV, 100 nA). The beams were operated in non-interlaced mode, alternating 1 analysis cycle and 1 frame of sputtering (corresponding to 1.31s) followed by a pause of 2s for the charge compensation. The $MCs_n^+$ ($n$ = 1, 2) depth profiling has also been used to improve the understanding of the data. This is a useful method, mainly applied to quantify the alloys but also to identify any ion compounds. The cesium primary beam is used for sputtering during the depth profile and permits the detection of $MCs^+$ or $MCs_2^+$ cluster ions where $M$ is the element of interest combined with one or two Cs atoms. The advantages of following $MCs^+$ and $MCs_2^+$ ions during ToF-SIMS analysis include reducing matrix effects and the possibility of detecting the compounds from both electronegative and electropositive elements and compounds. All depth profiles have been point-to-point normalized by the total ion intensity and the data have been plotted using a 5-points adjacent averaging. Both normalization and smoothing have permitted a better comparison of the data from the different samples. The depth calibrations have been established using the interface tool in SurfaceLab version 7.3 software from ION-TOF GmbH to identify the different interfaces and based on the measured thicknesses using the surface profiler to obtain a line scan of the craters with the in-situ SPM by contact scanning. Chemical mappings were obtained using a field of view of 20 × 20 µm$^2$, with a raster of 2048 by 2048 pixels and then the image raster has been pinned by a factor 64 to enhance the signal-to-noise ratio. Ion signals from ion mappings have been normalized using the total ion signal to standardize the values and to help for the comparison between the different samples or the area.

**Transport measurements**

All transport measurements were performed using a 450PM Manual Probe Station Micromanipulator. For 2-terminal IV measurements, 2 probes were dropped on a pair of Au electrodes, forming the ends of a closed circuit. For 3-terminal FET measurements, 3 probes were dropped over a pair of Au electrodes and the back contact. For temperature-dependent measurements, a custom-built setup comprising of a source meter and relay interface were used with a heating element, thermocouple, and copper plate for heat homogenization.

Mobility calculations were performed using peak transconductance ($\mu_{g_m}$) :

$$\mu_{g_m} = \frac{dI_{DS}}{dV_{GS}} \left(\frac{L_{CH}}{W \, C_{OX} \, V_{DS}}\right) \ldots (3)$$

Subthreshold-swing measurements were made from the linear regime of the $I_{DS}$-$V_{GS}$ plot, while following cautionary protocols mentioned elsewhere[63].

**DFT calculations**

First-principles calculations based on DFT[64] for adsorption energy were carried out using the plane-wave basis Vienna Ab initio Simulation Package[65,66]. The Perdew-Burke-Ernzerhof[67] version of generalized gradient approximation was chosen to describe the electron exchange-correlation interactions, with a cutoff energy of 500 eV. The Au (111) surface was modeled using a three-layer 6 × 6 supercell slab with the bottom two layers fixed. The SiO$_2$ surface was modeled using an oxygen-exposed 8 × 8 supercell slab with 1 × 2 reconstruction fully considered. The WSe$_2$ surface was modeled using a two-layer 5 × 5 supercell slab with the bottom monolayer WSe$_2$ fixed. The vacuum layer thickness was ensured to be larger than 15 Å to avoid spurious interactions between neighboring images. The van der Waals interactions were considered for the structure optimization using the DFT-D3 approach[68]. The structures were fully relaxed using a 1 × 1 $k$-mesh until the force on each unconstrained atom is less than 0.01 eV/Å. The adsorption energy was calculated as follows: $E_{adsorption} = E_{tot} - E_{substrate} - E_{WSe_6}$, where $E_{tot}$ denotes the total energy of the WSe$_6$ molecule/substrate system, $E_{substrate}$ is the total energy of the according surface, and $E_{WSe_6}$ is the total energy of WSe$_6$ molecule.

**Bandgap-strain and electron transport calculations**

Band gaps of WSe$_2$ under strain were calculated within DFT using the PBE exchange-correlation functional[67] and double-ζ plus polarization (DZP) basis set, as implemented in the SIESTA code[69]. At each strain state, the WSe$_2$ cell is kept fixed, while atomic positions are relaxed until the residual force on each atom is smaller than 0.01 eV/Ang. A 20✕20 *k*-point mesh is used.

The Au|WSe$_2$ interface electron transport properties are calculated using DFT and Non-Equilibrium Green's Function(NEGF) method as available in the TranSIESTA code[70]. Singe-ζ plus polarization (SZP) basis set is chosen due to high computational cost. 10✕15 *k*-point mesh and 1✕20 *k*-point mesh is used for electrodes and transport calculation, respectively.

**Contributions**

L.M.S. and S.A.I. contributed equally to this manuscript. L.M.S., S.A.I., A.B.P. and P.M.A. were involved in the design and conception of the experiments. L.M.S. and S.A.I. performed the growth, characterization, and fabrication of the in-situ devices. S.A.I. and A.B.P performed the device measurements. A.B.P. performed the TEM measurements and S.A.I performed the analysis. X.L., Y.H., and B.I.Y. designed the simulations. X.L. performed the DFT surface energy calculations, and Y.H. performed the ab initio strain simulations and IV curve simulations. T.T. performed the TOF-SIMS data analysis and treatment. A.M. and A.P.C.T. contributed to the optimization of samples. P.B., C.S.T., R.V., and S.T. contributed to the discussion and analysis of the results. The manuscript was written through the contributions of all authors.

**Acknowledgements**

The authors would like to greatly thank Prof. Venkataraman Swaminathan for his advice and guidance through intellectual discussions. The authors thank Prof. Junichiro Kono, Dr. Natsumi Komatsu, and Hongjing Xu for their assistance in setting up a heating stage for high-temperature IV measurements. S. A. I. and P. M. A. would like to acknowledge the US Air Force Research Labs and UES for financial support. ToF-SIMS analysis were carried out with support provided by the National Science Foundation CBET-1626418. L.M.S acknowledges CAPES (Coordination for the Improvement of Higher Education Personnel) under the Brazilian Ministry of Education for the financial support in the form of a graduate fellowship. Computational modeling work (Y.H., X.L. and B.I.Y.) was supported by the US Department of Energy, BES grant DE-SC0012547 (aspects of synthesis) and by the Army Research Office grant W911NF-16-1-0255 (electronics, transport). This work conducted in part using resources of the Shared Equipment Authority at Rice University.

**Competing Interest**

The authors S. A. I.; L. M. S.; A. B. P.; R. V.; and P. M. A. declare interest in pursuing relevant intellectual property: 'A Method to Prepare Tunable Schottky-Ohmic Contacts Between Metal and Two-Dimensional Materials, and Semiconductor Devices Thereof', Tech Id No. 2022-055.

**References**

1. Akinwande, D. *et al.* Graphene and two-dimensional materials for silicon technology. *Nature* **573**, 507–518 (2019).

2. Das, S. *et al.* Transistors based on two-dimensional materials for future integrated circuits. *Nat. Electron.* **4**, 786–799 (2021).

3. Tang, A. *et al.* Toward Low-Temperature Solid-Source Synthesis of Monolayer $MoS_2$. *ACS Appl. Mater. Interfaces* **13**, 41866–41874 (2021).

4.  Huh, W., Lee, D. & Lee, C. Memristors Based on 2D Materials as an Artificial Synapse for Neuromorphic Electronics. *Adv. Mater.* **32**, 2002092 (2020).

5.  Mennel, L. *et al.* Ultrafast machine vision with 2D material neural network image sensors. *Nature* **579**, 62–66 (2020).

6.  Liu, E. *et al.* Integrated digital inverters based on two-dimensional anisotropic ReS2 field-effect transistors. *Nat. Commun.* **6**, 6991 (2015).

7.  Chhowalla, M., Jena, D. & Zhang, H. Two-dimensional semiconductors for transistors. *Nat. Rev. Mater.* **1**, 16052 (2016).

8.  Iyengar, S. A., Puthirath, A. B. & Swaminathan, V. Realizing Quantum Technologies in Nanomaterials and Nanoscience. *Adv. Mater.* 2107839 (2022) doi:10.1002/adma.202107839.

9.  Abuzaid, H., Williams, N. X. & Franklin, A. D. How good are 2D transistors? An application-specific benchmarking study. *Appl. Phys. Lett.* **118**, 030501 (2021).

10. Fenouillet-Beranger, C. *et al.* Guidelines for intermediate back end of line (BEOL) for 3D sequential integration. in *2017 47th European Solid-State Device Research Conference (ESSDERC)* 252–255 (IEEE, 2017). doi:10.1109/ESSDERC.2017.8066639.

11. Fenouillet-Beranger, C. *et al.* A Review of Low Temperature Process Modules Leading Up to the First (≤500 °C) Planar FDSOI CMOS Devices for 3-D Sequential Integration. *IEEE Trans. Electron Devices* **68**, 3142–3148 (2021).

12. Watson, A. J., Lu, W., Guimarães, M. H. D. & Stöhr, M. Transfer of large-scale two-dimensional semiconductors: challenges and developments. *2D Mater.* **8**, 032001 (2021).

13. Huang, Y. *et al.* Reliable Exfoliation of Large-Area High-Quality Flakes of Graphene and Other Two-Dimensional Materials. *ACS Nano* **9**, 10612–10620 (2015).

14. Huang, Y. *et al.* Universal mechanical exfoliation of large-area 2D crystals. *Nat. Commun.* **11**, 2453 (2020).

15. Nicolosi, V., Chhowalla, M., Kanatzidis, M. G., Strano, M. S. & Coleman, J. N. Liquid Exfoliation of Layered Materials. *Science* **340**, 1226419 (2013).

16. Lee, Y.-H. *et al.* Synthesis of Large-Area $MoS_2$ Atomic Layers with Chemical Vapor Deposition. *Adv. Mater.* **24**, 2320–2325 (2012).

17. Zhan, Y., Liu, Z., Najmaei, S., Ajayan, P. M. & Lou, J. Large-Area Vapor-Phase Growth and Characterization of $MoS_2$ Atomic Layers on a $SiO_2$ Substrate. *Small* **8**, 966–971 (2012).

18. Kang, K. *et al.* High-mobility three-atom-thick semiconducting films with wafer-scale homogeneity. *Nature* **520**, 656–660 (2015).

19. Yue, R. *et al.* Nucleation and growth of $WSe_2$: enabling large grain transition metal dichalcogenides. *2D Mater.* **4**, 045019 (2017).

20. Kozhakhmetov, A., Torsi, R., Chen, C. Y. & Robinson, J. A. Scalable low-temperature synthesis of two-dimensional materials beyond graphene. *J. Phys. Mater.* **4**, 012001 (2020).

21. Zhou, J. *et al.* A library of atomically thin metal chalcogenides. *Nature* **556**, 355–359 (2018).

22. Song, I. *et al.* Patternable Large-Scale Molybdenium Disulfide Atomic Layers Grown by Gold-Assisted Chemical Vapor Deposition. *Angew. Chem. Int. Ed.* **53**, 1266–1269 (2014).

23. Wang, Z. *et al.* Metal Induced Growth of Transition Metal Dichalcogenides at Controlled Locations. *Sci. Rep.* **6**, 38394 (2016).

24. Chiu, M. *et al.* Metal-Guided Selective Growth of 2D Materials: Demonstration of a Bottom-Up CMOS Inverter. *Adv. Mater.* **31**, 1900861 (2019).

25. Parto, K. *et al.* One-Dimensional Edge Contacts to Two-Dimensional Transition-Metal Dichalcogenides: Uncovering the Role of Schottky-Barrier Anisotropy in Charge Transport across $MoS_2$/Metal Interfaces. *Phys. Rev. Appl.* **15**, 064068 (2021).

26. Zhang, D., Yeh, C.-H., Cao, W. & Banerjee, K. 0.5T0.5R—An Ultracompact RRAM Cell Uniquely Enabled by van der Waals Heterostructures. *IEEE Trans. Electron Devices* **68**, 2033–2040 (2021).

27. Li, S. *et al.* Halide-assisted atmospheric pressure growth of large WSe2 and WS2 monolayer crystals. *Appl. Mater. Today* **1**, 60–66 (2015).

28. Gao, Y. *et al.* Large-area synthesis of high-quality and uniform monolayer WS2 on reusable Au foils. *Nat. Commun.* **6**, 8569 (2015).

29. Gao, Y. *et al.* Ultrafast Growth of High-Quality Monolayer $WSe_2$ on Au. *Adv. Mater.* **29**, 1700990 (2017).

30. Yun, S. J. *et al.* Synthesis of Centimeter-Scale Monolayer Tungsten Disulfide Film on Gold Foils. *ACS Nano* **9**, 5510–5519 (2015).

31. Tonndorf, P. *et al.* Photoluminescence emission and Raman response of monolayer MoS2, MoSe2, and WSe2. *Opt Express* **21**, 4908–4916 (2013).

32. Thompson, C. V. Solid-State Dewetting of Thin Films. *Annu. Rev. Mater. Res.* **42**, 399–434 (2012).

33. Leroy, F. *et al.* How to control solid state dewetting: A short review. *Surf. Sci. Rep.* **71**, 391–409 (2016).

34. Zhang, R., Drysdale, D., Koutsos, V. & Cheung, R. Controlled Layer Thinning and p-Type Doping of $WSe_2$ by Vapor $XeF_2$. *Adv. Funct. Mater.* **27**, 1702455 (2017).

35. Ruano, G. *et al.* Stages of Se adsorption on Au(111): A combined XPS, LEED, TOF-DRS, and DFT study. *Surf. Sci.* **662**, 113–122 (2017).

36. Shenasa, M., Sainkar, S. & Lichtman, D. XPS study of some selected selenium compounds. *J. Electron Spectrosc. Relat. Phenom.* **40**, 329–337 (1986).

37. Dhayagude, A. C., Maiti, N., Debnath, A. K., Joshi, S. S. & Kapoor, S. Metal nanoparticle catalyzed charge rearrangement in selenourea probed by surface-enhanced Raman scattering. *RSC Adv.* **6**, 17405–17414 (2016).

38. Choi, M. S. *et al.* High carrier mobility in graphene doped using a monolayer of tungsten oxyselenide. *Nat. Electron.* **4**, 731–739 (2021).

39. Li, S. *et al.* Vapour–liquid–solid growth of monolayer MoS2 nanoribbons. *Nat. Mater.* **17**, 535–542 (2018).

40. Rasouli, H. R., Mehmood, N., Çakıroğlu, O. & Kasırga, T. S. Real time optical observation and control of atomically thin transition metal dichalcogenide synthesis. *Nanoscale* **11**, 7317–7323 (2019).

41. Lei, J., Xie, Y. & Yakobson, B. I. Gas-Phase "Prehistory" and Molecular Precursors in Monolayer Metal Dichalcogenides Synthesis: The Case of $MoS_2$. *ACS Nano* **15**, 10525–10531 (2021).

42. Lei, J., Xie, Y., Kutana, A., Bets, K. V. & Yakobson, B. I. Salt-Assisted $MoS_2$ Growth: Molecular Mechanisms from the First Principles. *J. Am. Chem. Soc.* **144**, 7497–7503 (2022).

43. Zavabeti, A. *et al.* A liquid metal reaction environment for the room-temperature synthesis of atomically thin metal oxides. *Science* **358**, 332–335 (2017).

44. Sutter, P., French, J. S., Khosravi Khorashad, L., Argyropoulos, C. & Sutter, E. Optoelectronics and Nanophotonics of Vapor–Liquid–Solid Grown GaSe van der Waals Nanoribbons. *Nano Lett.* **21**, 4335–4342 (2021).

45. Li, X. *et al.* Nickel particle–enabled width-controlled growth of bilayer molybdenum disulfide nanoribbons. *Sci. Adv.* **7**, eabk1892 (2021).

46. Liu, J., Fan, X., Shi, Y., Singh, D. J. & Zheng, W. Melting of Nanocrystalline Gold. *J. Phys. Chem. C* **123**, 907–914 (2019).

47. Graf, N., Gross, T., Wirth, T., Weigel, W. & Unger, W. E. S. Application of XPS and ToF-SIMS for surface chemical analysis of DNA microarrays and their substrates. *Anal. Bioanal. Chem.* **393**, 1907–1912 (2009).

48. Ahn, G. H. *et al.* Strain-engineered growth of two-dimensional materials. *Nat. Commun.* **8**, 608 (2017).

49. Bhanu, U., Islam, M. R., Tetard, L. & Khondaker, S. I. Photoluminescence quenching in gold - MoS2 hybrid nanoflakes. *Sci. Rep.* **4**, 5575 (2014).

50. Yang, B. *et al.* Effect of Distance on Photoluminescence Quenching and Proximity-Induced Spin–Orbit Coupling in Graphene/WSe$_2$ Heterostructures. *Nano Lett.* **18**, 3580–3585 (2018).

51. Zhang, W. *et al.* Paraffin-Enabled Compressive Folding of Two-Dimensional Materials with Controllable Broadening of the Electronic Band Gap. *ACS Appl. Mater. Interfaces* **13**, 40922–40931 (2021).

52. Li, H. *et al.* Anomalous Behavior of 2D Janus Excitonic Layers under Extreme Pressures. *Adv. Mater.* **32**, 2002401 (2020).

53. Liu, D. *et al.* Substrate effect on the photoluminescence of chemical vapor deposition transferred monolayer WSe$_2$. *J. Appl. Phys.* **128**, 043101 (2020).

54. Gusakova, J. *et al.* Electronic Properties of Bulk and Monolayer TMDs: Theoretical Study Within DFT Framework (GVJ-2e Method). *Phys. Status Solidi A* **214**, 1700218 (2017).

55. DiStefano, J. G., Murthy, A. A., Jung, H. J., dos Reis, R. & Dravid, V. P. Structural defects in transition metal dichalcogenide core-shell architectures. *Appl. Phys. Lett.* **118**, 223103 (2021).

56. Hannon, J. B., Kodambaka, S., Ross, F. M. & Tromp, R. M. The influence of the surface migration of gold on the growth of silicon nanowires. *Nature* **440**, 69–71 (2006).

57. Liu, H., Neal, A. T., Zhu, Z., Tománek, D. & Ye, P. D. Phosphorene: A New 2D Material with High Carrier Mobility.

58. Chen, R.-S., Tang, C.-C., Shen, W.-C. & Huang, Y.-S. Thickness-dependent electrical conductivities and ohmic contacts in transition metal dichalcogenides multilayers. *Nanotechnology* **25**, 415706 (2014).

59. Aboelfotoh, M. O. Temperature dependence of the Schottky-barrier height of tungsten on n-type and p-type silicon. *Solid-State Electron.* **34**, 51–55 (1991).

60. Liu, Y. *et al.* Approaching the Schottky–Mott limit in van der Waals metal–semiconductor junctions. *Nature* **557**, 696–700 (2018).

61. Sebastian, A., Pendurthi, R., Choudhury, T. H., Redwing, J. M. & Das, S. Benchmarking monolayer MoS2 and WS2 field-effect transistors. *Nat. Commun.* **12**, 693 (2021).

62. Datye, I. M. *et al.* Strain-Enhanced Mobility of Monolayer $MoS_2$. *Nano Lett.* **22**, 8052–8059 (2022).

63. McCulloch, I., Salleo, A. & Chabinyc, M. Avoid the kinks when measuring mobility. *Science* **352**, 1521–1522 (2016).

64. Kohn, W. & Sham, L. J. Self-Consistent Equations Including Exchange and Correlation Effects. *Phys. Rev.* **140**, A1133–A1138 (1965).

65. Kresse, G. & Furthmüller, J. Efficient iterative schemes for *ab initio* total-energy calculations using a plane-wave basis set. *Phys. Rev. B* **54**, 11169–11186 (1996).

66. Kresse, G. & Joubert, D. From ultrasoft pseudopotentials to the projector augmented-wave method. *Phys. Rev. B* **59**, 1758–1775 (1999).

67. Perdew, J. P., Burke, K. & Ernzerhof, M. Generalized Gradient Approximation Made Simple. *Phys. Rev. Lett.* **77**, 3865–3868 (1996).

68. Grimme, S., Antony, J., Ehrlich, S. & Krieg, H. A consistent and accurate *ab initio* parametrization of density functional dispersion correction (DFT-D) for the 94 elements H-Pu. *J. Chem. Phys.* **132**, 154104 (2010).

69. Soler, J. M. *et al.* The SIESTA method for ab initio order-N materials simulation. *J. Phys. Condens. Matter* **14**, 2745 (2002).

70. Brandbyge, M., Mozos, J.-L., Ordejón, P., Taylor, J. & Stokbro, K. Density-functional method for nonequilibrium electron transport. *Phys. Rev. B* **65**, 165401 (2002).